\documentclass[%
 reprint,
 superscriptaddress,
 amsmath,amssymb,
 aps,
]{revtex4-2}

\usepackage{graphicx}
\usepackage{dcolumn}
\usepackage{bm}
\usepackage{hyperref}

\begin{document}

\title{Upper bound inequality for calculation time in simulated annealing analogous to adiabatic theorem in quantum systems}

\author{Akihisa Ichiki}
\email{ichiki@chem.material.nagoya-u.ac.jp}
\affiliation{Institutes of Innovation for Future Society, Nagoya University, Furo-cho, Chikusa-ku, Nagoya 464-8603, Japan}
\author{Masayuki Ohzeki}
\affiliation{Graduate School of Information Sciences, Tohoku University, Sendai 980-8579, Japan}
\affiliation{Institute of Innovative Research, Tokyo Institute of Technology, Oh-okayama, Meguro-ku, Tokyo 152-8550, Japan}
\affiliation{Sigma-i, Co. Ltd., Konan, Minato-ku, Tokyo 108-0075, Japan}
\date{\today}

\begin{abstract}
It has been recently reported that classical systems have speed limit for state evolution, although such a concept of speed limit had been considered to be unique to quantum systems. 
Owing to the speed limit for classical system, the lower bound for calculation time of simulated annealing with desired calculation accuracy can be derived. 
However, such a lower bound does not work as a criterion for completion of calculation in a practical time. 
In this paper, we derive an inequality for classical system analogous to the quantum adiabatic theorem that gives calculation time for an accuracy-guaranteed fluctuation-exploiting computation.
The trade-off relation between calculation time and accuracy is given in the form tolerable in practical use.
\end{abstract}

\maketitle

\section{Introduction}

Exploiting fluctuation has become an indispensable technique to solve optimization problems both in classical and quantum computations, i.e., simulated~\cite{Kirkpatrick1983, Cerny1985, Otten1989} and quantum annealing~\cite{PhysRevE.58.5355, 1994CPL...219..343F, Farhi472, Santoro2427, RevModPhys.80.1061, Johnson2011Quantum}. 
The original optimization problem is appropriately mapped to a spin model \cite{10.3389/fphy.2014.00005}. 
Then the optimum solution for the original problem is translated as a ground state of the spin system that is realized through its natural relaxation process. 
In such a general calculation scheme, relaxation time of the system is regarded as the calculation time to solve the optimization problem. 

The relaxation of the quantum system is governed by the so-called quantum speed limit (QSL)~\cite{Mandelstam1991, MARGOLUS1998188, Lloyd2000, PhysRevA.67.052109, UHLMANN1992329, PhysRevLett.70.3365, Deffner_2013, PhysRevLett.110.050402, PhysRevLett.110.050403, PhysRevLett.111.010402}. 
QSL gives the lower bound of the transition time of quantum systems from given initial state to the provided final state. 
QSL yields an uncertainty relation between the time of transition and the energy gap between two states. 
It has been reported that classical systems also have such a concept of speed limit, namely classical speed limit (CSL)~\cite{PhysRevLett.120.070401, PhysRevLett.120.070402}, although such an uncertainty relation had been considered to be unique in quantum systems. 
Then the lower bound of the relaxation time in classical system is roughly evaluated by CSL. 
The CSL is important for evaluating the limit of the calculation speed in principle. 
Unlike simulated annealing where Hamiltonian is driven externally, CSL provides the limit on the relaxation speed of the system under a fixed Hamiltonian. 
However, CSL allows us to estimate the speed limit for simulated annealing to approach a solution. 
The lower limit of the calculation time given by CSL is roughly determined only by the desired calculation accuracy and initial state, and does not depend on the details of the system Hamiltonian. Therefore, CSL imposes a computational speed limit that cannot be achieved by any arrangement of the Hamiltonian.

On the other hand, in order to judge whether the relaxation of the classical system is tolerable for practical use to solve optimization problems, it is required to evaluate the upper bound of the relaxation time rather than the lower bound, which is given by the CSL. 
In this paper, we will obtain such an upper bound in an inequality form. 
The derived inequality is regarded as the classical version of adiabatic theorem in quantum systems~\cite{1928ZPhy...51..165B, 1950JPSJ....5..435K}. 
In the context of fluctuation-exploiting computation, the adiabatic theorem implies that the calculation time is inversely proportional to the calculation accuracy. 
The coefficient in such a trade-off relation depends on the energy gap between the ground and the first exited states, and the transition amplitude between these two states. 
Because of the difficulty to concretely evaluate the transition amplitude, it is difficult to evaluate the relaxation time exactly from adiabatic theorem. 
On the other hand, our inequality has a simple form to easily evaluate the upper bound depending only on the initial state and the energy gap. 
Therefore, the calculation time in the worst case can be discussed focusing only on the energy gap without being bothered by estimating the transition amplitude in our framework. 


\section{Classical speed limit}
CSL was given independently in \cite{PhysRevLett.120.070401} and \cite{PhysRevLett.120.070402}. Shanahan, Chenu, Margolus, and del Campo derived the CSL as an uncertainty relation by phase-space approach in \cite{PhysRevLett.120.070401}. On the other hand, Okuyama and Ohzeki gave the CSL directly from stochastic dynamics in \cite{PhysRevLett.120.070402}. 
They consider the Fokker-Planck equation
\begin{eqnarray}
\dfrac{\partial}{\partial t}\rho(x, t) = \dfrac{\partial}{\partial x}\left[2\dfrac{\partial W(x)}{\partial x} + \dfrac{\partial}{\partial x}\right]\rho(x, t)\,,
\end{eqnarray}
which has a steady state solution $\pi(x) = \exp\left[-2W(x)\right]$. 
Rewriting $\rho(x, t)$ as $\rho(x, t) = \exp\left[-W(x)\right]\psi(x, t)$, we have the imaginary-time Schr\"odinger equation as 
\begin{eqnarray}
-\dfrac{\partial}{\partial t}\psi(x, t) &=& \left[-\dfrac{\partial^2}{\partial x^2} + \left(\dfrac{\partial W}{\partial x}\right)^2 - \dfrac{\partial^2 W}{\partial x^2}\right]\psi(x, t)\nonumber\\
&:=& \hat{H}_{F}\psi(x, t)\,.
\end{eqnarray} 
The ground state for $\hat{H}_F$ is given by $\psi_0(x) = \exp\left[-W(x)\right]$. 
Then the CSL corresponding to the QSL known as the Margolus-Levitin bound~\cite{MARGOLUS1998188} is given as 
\begin{eqnarray}
\tau \ge \dfrac{\ln\left\langle\psi(0)|\psi(0)\right\rangle - \ln\left\langle\psi(0)|\psi(\tau)\right\rangle}{\dfrac{\left\langle\psi(0)|\hat{H}_{F}|\psi(0)\right\rangle}{\left\langle\psi(0)|\psi(0)\right\rangle}} := \tau_{\rm min}\,.\label{min}
\end{eqnarray}
Note that the overlap between the initial and final states, $\left\langle\psi(0)|\psi(\tau)\right\rangle$, characterizes the deviation of the final state $\psi(\tau)$ from the ground state $\psi_0$, which is regarded as the calculation error in the context of simulated annealing. 
Note also that the $\tau_{\rm min}$ gives the lower limit of relaxation time for the time-independent Fokker-Planck operator. 
Thus $\tau_{\rm min}$ should be used as a guideline for the calculation speed limit, not as the calculation speed limit itself, in the use of simulated annealing where the Fokker-Planck operator varies temporally. 
Eq.~(\ref{min}) implies that, in the context of simulated annealing, the minimum required time to solve the optimization problem is roughly bounded by $\tau_{\rm min}$ that is evaluated only by desired calculation accuracy and initial condition. 
However, such a lower bound of calculation time cannot be a criterion of completion of simulated annealing in a practical time, since $\tau_{\rm min}$ underestimates the calculation time. 
In section~\ref{main_sec}, on the contrary, we will derive the upper bound of the calculation time that gives a criterion for a practical use of simulated annealing.

\section{Adiabatic theorem}

In this section, we will overview the quantum adiabatic theorem, which provides the trade-off relation between calculation time and accuracy in quantum annealing. 

Consider the temporally modified system Hamiltonian 
\begin{eqnarray}
\hat{H}(t) = \dfrac{t}{\tau}\hat{H}_0 + \left(1-\dfrac{t}{\tau}\right)\hat{H}_1\,,
\end{eqnarray}
where $\hat{H}_0$ is the target Hamiltonian corresponding to the considered optimization problem and $\hat{H}_1$ is an initial Hamiltonian whose ground state can be easily prepared by some operations. 
In our system settings, the temporal evolution of the system state, which is regarded as the calculation process to solve the optimization problem, is assumed to be stopped when the system Hamiltonian $\hat{H}(t)$ becomes the target Hamiltonian $\hat{H}_0$.  The calculation time $\tau$, which can be an arbitrary positive value, is given by the protocol of the system Hamiltonian. 
The calculation error $\delta$ is characterized by the overlap between the ground state $|\psi_0(\tau)\rangle$ of the target Hamiltonian $\hat{H}_0$ and the system state $|\psi(\tau)\rangle$ at the final time as 
\begin{eqnarray}
\left| \left\langle\psi(\tau)|\psi_0(\tau)\right\rangle\right|^2 = 1-\delta^2\,.
\end{eqnarray}
We assume here that the error is sufficiently small, i.e., $\delta^2 \ll 1$. 
The calculation time $\tau$ is related to $\delta$ as 
\begin{eqnarray}
\dfrac{\max_t \left\langle\psi_1(t)|\hat{H}_0-\hat{H}_1|\psi_0(t)\right\rangle}{\tau\min_t \Delta_t^2} = \delta\,,\label{QAT}
\end{eqnarray}
where $|\psi_0(t)\rangle$ and $|\psi_1(t)\rangle$ denote the ground and the first excited states for the Hamiltonian $\hat{H}(t)$ at time $t$, respectively, and $\Delta_t$ is the energy gap between these two states \cite{doi:10.1063/1.2995837}. The maximization and minimization are carried out over time $0 \le t \le \tau$. 
Thus the trade-off between the calculation time and accuracy is roughly given as 
\begin{eqnarray}
\tau\propto\dfrac{1}{\delta\min_t\Delta_t^2}\,.
\end{eqnarray}
This is the quantum adiabatic theorem, which implies that longer calculation time leads less error, and greater energy gap is preferable both for less error and shorter calculation time. 
However, it is difficult to concretely evaluate the calculation time with required accuracy from Eq.~(\ref{QAT}) because of the transition amplitude from the ground to the first excited state in Eq.~(\ref{QAT}). 
In the next section, we will derive the simple inequality that can be regarded as the relaxed version of Eq.~(\ref{QAT}) for classical systems.

\section{upper bound inequality}\label{main_sec}

In this section, we will investigate the upper bound of calculation time for simulated annealing, which is realized in classical systems, in contrast to the lower bound given by CSL. 

Consider a Fokker-Planck equation 
\begin{eqnarray}
\dfrac{\partial \rho(t)}{\partial t} = -\hat{L}_{\gamma(t), u(t)} \rho(t)\,,\label{FPE}
\end{eqnarray}
where the Fokker-Planck operator $\hat{L}_{\gamma(t), u(t)}$ depends on time via parameters $\gamma$ and $u$ which are controlled during the calculation process. 
The parameter $\gamma$ controls the quantities such as noise strength, and it does not violate the detailed balance condition. 
On the contrary, the parameter $u$ yields a probability flow, which characterizes the violation of detailed balance condition \cite{PhysRevE.88.020101, PhysRevE.91.062105}.  
Hereinafter, we assume that the Fokker-Planck operator $\hat{L}_{\gamma(t), u(t)}$ has a unique steady state solution for each $t$. 
The Fokker-Planck operator $\hat{L}_{\gamma, u}$ is assumed to be driven externally by the temporal changes of $\gamma$ and $u$. 
The initial state is set to be the steady state for the operator $\hat{L}_{\gamma(0), u(0)}$, which is easily prepared. 
The steady state for the final operator $\hat{L}_{\gamma(\tau), u(\tau)}$ is set to correspond to the optimum solution for the considered problem. 
The Fokker-Planck operator $\hat{L}_{\gamma(t), u(t)}$ is characterized by its eigenvalues and eigenfunctions as follows: 
\begin{eqnarray}
\hat{L}_{\gamma(t), u(t)}\phi_k^{\gamma(t), u(t)} &=& \lambda_k^{\gamma(t), u(t)} \phi_k^{\gamma(t), u(t)}\,,\\
\hat{L}_{\gamma(t), u(t)}^\dagger(t)\psi_k^{\gamma(t), u(t)} &=& \lambda_k^{\gamma(t), u(t)} \psi_k^{\gamma(t), u(t)}\,,\label{adjoint}
\end{eqnarray}
where $\hat{L}_{\gamma, u}^\dagger$ indicates an adjoint operator of the Fokker-Planck operator. 
The eigenvalues are ordered as $0 = \lambda_0^{\gamma(t), u(t)} \le{\rm Re}\lambda_1^{\gamma(t), u(t)}\le{\rm Re}\lambda_2^{\gamma(t), u(t)}\le\cdots$. 
The eigenfunctions satisfy the orthonormal relations: 
\begin{eqnarray}
\int dx\, \psi_k^{\gamma(t), u(t)}(x)\phi_l^{\gamma(t), u(t)}(x) = \delta_{kl}\,,\label{ortho}
\end{eqnarray}
where $\delta_{kl}$ is a Kronecker delta and 
\begin{eqnarray}
\displaystyle\sum_n \psi_n^{\gamma(t), u(t)}(x) \phi_n^{\gamma(t), u(t)}(y) = \delta(x - y)\,.\label{delta}
\end{eqnarray}
When the detailed balance condition holds, i.e., $u = 0$, the eigenvalues are all real and the eigenfunctions satisfies 
\begin{eqnarray}
\phi_k^{\gamma(t), 0} = \psi_k^{\gamma(t), 0} \pi^{\gamma(t)}\,,\label{eigenDBC}
\end{eqnarray}
where $\pi^{\gamma(t)}$ is the steady state solution $\hat{L}_{\gamma(t), 0} \pi^{\gamma(t)} = 0$, and $\pi^{\gamma(t)} = \phi_0^{\gamma(t), 0}$ by definition of eigenfunctions. 
Eq.~(\ref{eigenDBC}) is straightforwardly shown by substituting it into the characteristic equation~(\ref{adjoint}) for the adjoint operator $\hat{L}_{\gamma(t), u(t)}^\dagger$ with $u(t) = 0$.  
Note that the steady state $\pi^{\gamma(t)}$ is independent of the value of $u$ which controls a steady state probability flow, i.e., $\hat{L}_{\gamma(t), u(t)}\pi^{\gamma(t)} = 0$ for arbitrary $u(t)$ \cite{PhysRevE.88.020101}. 
For convenience, the steady state solution is normalized as $\int dx\,\pi^{\gamma(t)} = 1$ for arbitrary $t$, where the integration is carried out over all system degrees of freedom. 
In the absence of detailed balance condition where $u(t)\neq 0$, eigenvalues are complex and the simple relation (\ref{eigenDBC}) does not hold in general. 

Consider the expansion of the probability density $\rho(t)$, which follows the Fokker-Planck equation (\ref{FPE}), in terms of the eigenfunctions at the final time in the presence of detailed balance condition, i.e., the eigenfunctions for $\hat{L}_{\gamma(\tau), 0}$ as 
\begin{eqnarray}
\rho(t) = \displaystyle\sum_n c_n(t)\phi_n^{\gamma(\tau), 0}\,.\label{expansion}
\end{eqnarray}
By our assumption that $\hat{L}_{\gamma(t), u(t)}$ has a unique stationary solution for fixed $t$, an arbitrary state $\rho(t)$ converges to the stationary solution $\pi^{\gamma(t)}$ under the dynamics with fixed Fokker-Planck operator $\hat{L}_{\gamma(t), u(t)}$ after a long time.  
Then in addition to Eq.~(\ref{expansion}), we consider the expansion of $\rho(t)$ by the eigenfunctions for $\hat{L}_{\gamma(t), u(t)}$, which give orthonormal basis varying temporally: 
\begin{eqnarray}
\rho(t) = \displaystyle\sum_n d_n(t)\phi_n^{\gamma(t), u(t)}\,.\label{expansion2}
\end{eqnarray}
Since the steady state solution $\phi_0^{\gamma(t), u(t)} = \pi^{\gamma(t)}$ and $\rho(t)$ both are normalized as $\int dx\, \pi^{\gamma(t)} = 1$ and $\int dx\,\rho(t) = 1$, we find $d_n(t) = 0$ and $\int dx\,\phi_n^{\gamma(t), u(t)} = 0$ for $n\ge 1$.  
The relation $c_0(t)=1$ is also easily derived from Eq.~(\ref{expansion}).   
Comparing the two expression for an arbitrary distribution Eqs.~(\ref{expansion}) and (\ref{expansion2}), we find the coefficient $d_m(t)$ as 
\begin{eqnarray}
d_m(t) = \displaystyle\sum_n c_n(t) A_{nm}(t)\,,\label{b}
\end{eqnarray}
where the matrix $A(t)$ is defined as 
\begin{eqnarray}
A_{nm}(t) = \int dx\,\psi_m^{\gamma(t), u(t)}\phi_n^{\gamma(\tau), 0}\,.\label{A}
\end{eqnarray}
Furthermore, it is straightforwardly shown that the inverse of the matrix $A(t)$ is given as 
\begin{eqnarray}
\left(A^{-1}(t)\right)_{nm} = \int dx\psi_m^{\gamma(\tau), 0}\phi_n^{\gamma(t), u(t)}\label{invA}
\end{eqnarray}
by using the orthonormal relations Eqs.~(\ref{ortho}) and (\ref{delta}).  
Substituting the expression~(\ref{expansion}) into the left-hand-side and Eq.~(\ref{expansion2}) into the right-hand-side of the Fokker-Planck equation~(\ref{FPE}) respectively, and using Eqs.~(\ref{b})--(\ref{invA}), we obtain the dynamics for $c_n(t)$ as 
\begin{eqnarray}
\dot{c}_n(t) = -\displaystyle\sum_{m, k}c_m(t) A_{mk}(t)\lambda_k^{\gamma(t), u(t)}A_{kn}^{-1}(t)\,.\label{cdynamics}
\end{eqnarray}

In order to evaluate the calculation error quantitatively, we consider the following overlap between probability densities: 
\begin{eqnarray}
D(\tau) = \int\dfrac{dx}{\pi^{\gamma(\tau)}}\rho(0)\rho(\tau)\,.
\end{eqnarray}
Since $\phi_k^{\gamma(\tau), 0} = \psi_k^{\gamma(\tau), 0}\pi^{\gamma(\tau)}$, this overlap is essentially same as the overlap $\left\langle\psi(0)|\psi(\tau)\right\rangle$ appearing in Eq.~(\ref{min}). 
Using the expression of the probability density Eq.~(\ref{expansion}) and the fact that $c_0(t) = 1$ for arbitrary $t$, the overlap is rewritten as 
\begin{eqnarray}
D(\tau) = 1 + \vec{c}(0)^{\rm T}\vec{c}(\tau)\,,
\end{eqnarray}
where $\vec{c}(t) = \left(c_1(t), c_2(t), \cdots\right)^{\rm T}$. 
Here, we have used the relation~(\ref{eigenDBC}) for $\phi_n^{\gamma(\tau), 0}$ and the orthonormal condition~(\ref{ortho}).  
According to Schwartz inequality, we obtain 
\begin{eqnarray}
D(\tau) \le 1 + \left|\vec{c}(0)\right|\left|\vec{c}(\tau)\right|\,.\label{Schwartz}
\end{eqnarray}
Then $\left|\vec{c}(\tau)\right|$ is evaluated as follows: in the right-hand-side of Eq.~(\ref{cdynamics}), $A\Lambda A^{-1}$ gives a similarity transformation of $\Lambda = {\rm diag}(\lambda_0^{\gamma, u}, \lambda_1^{\gamma, u},\cdots)$. Thus the eigenvalues for $A\Lambda A^{-1}$ are same as those for $\Lambda$. 
Furthermore, considering the order of the eigenvalues $0=\lambda_0^{\gamma(t), u(t)}\le{\rm Re}\lambda_1^{\gamma(t), u(t)}\le{\rm Re}\lambda_2^{\gamma(t), u(t)}\le\cdots$ and the fact that $\lambda_n^{\gamma(t), u(t)}$ changes its value temporally, we have 
\begin{eqnarray}
\left|\vec{c}(\tau)\right| \le \exp\left[-\tau\displaystyle\min_t {\rm Re}\lambda_1^{\gamma(t), u(t)}\right]\left| \vec{c}(0)\right|\,.\label{eigenineq}
\end{eqnarray}
Using Eqs.~(\ref{Schwartz}) and (\ref{eigenineq}), we conclude 
\begin{eqnarray}
D(\tau) \le 1 + \left[D(0)-1\right]\exp\left[-\tau\displaystyle\min_t{\rm Re}\lambda_1^{\gamma(t), u(t)}\right]\,.
\end{eqnarray}
This is equivalent to the expression for the calculation time $\tau$ as 
\begin{eqnarray}
\tau \le \dfrac{1}{\min_t {\rm Re}\lambda_1^{\gamma(t), u(t)}}\ln\left[\dfrac{D(0)-1}{D(\tau)-1}\right] := \tau_{\rm max}^{\gamma, u}\,.\label{main}
\end{eqnarray}
Note that $D(\tau) = 1$ when $\rho(\tau) = \pi^{\gamma(\tau)}$, and $\pi^{\gamma(\tau)}$ is the steady state corresponding to the true solution of the considered optimization problem. 
Thus $D(\tau) - 1$ in the right-hand-side of Eq.~(\ref{main}) is regarded as a calculation error after $\tau$. 
Moreover, $D(0) - 1$ is determined only by the initial state, and ${\rm Re}\lambda_1^{\gamma(t), u(t)}$ plays the role of energy gap between the ground and the first excited states in the case of quantum annealing. 
Thus the inequality~(\ref{main}) is regarded as the classical version of adiabatic theorem~(\ref{QAT}). 
The inequality (\ref{main}) indicates that the calculation time of $\tau_{\rm max}^{\gamma, u}$ is sufficient to guarantee the calculation error less than $D(\tau)-1$. 
In contrast to the quantum adiabatic theorem~(\ref{QAT}), our bound is loose, but avoid the difficulty of the estimation of transition amplitudes.

\section{Summary and Discussion}

The inequality we have obtained gives the worst evaluation, while the CSL gives the best evaluation for the calculation time and accuracy. In practice, the worst evaluation for the calculation time and accuracy should be used as a criterion for completion of the calculation in practical time. 
The lower limit of calculation time $\tau_{\rm min}$ given by the CSL cannot be used for guaranteeing the calculation accuracy. For example, even if $\tau_{\rm min}$ is one second, the actual calculation time to achieve the desired calculation accuracy may be one month. On the other hand, the $\tau_{\rm max}^{\gamma, u}$ obtained by our result guarantees the calculation accuracy. If $\tau_{\rm max}^{\gamma, u}$ is ten second, the desired calculation accuracy is always achieved within ten second.

Inequality~(\ref{main}) gives the upper bound of the calculation time for simulated annealing in a simple form. 
Particularly, if the calculation error $D(\tau) - 1$ is preset as a target value, $\ln\left[\left(D(0)-1\right)/\left(D(\tau)-1\right)\right]$ in the right-hand-side does not depend on the relaxation dynamics. 
${\rm Re}\lambda_1^{\gamma(t), u(t)}$ only depends on the dynamics of the system. 
Thus, in order to shorten the calculation time, it is required to focus on a good design of annealing dynamics for greater ${\rm Re}\lambda_1^{\gamma, u}$.  

In the context of fasten convergence of Markov chain Monte Carlo algorithms~\cite{PhysRevLett.58.86, doi:10.1143/JPSJ.65.1604, Neal2001, PhysRevLett.105.120603, TURITSYN2011410, FERNANDES20111856, doi:10.7566/JPSJ.82.064003}, it is known that the violation of detailed balance yields the shift of the real part of the eigenvalues for the Fokker-Planck operator~\cite{PhysRevE.88.020101, PhysRevE.92.012105, Ohzeki_2015}: 
\begin{eqnarray}
{\rm Re}\lambda_1^{\gamma, u} \ge \lambda_1^{\gamma, 0}\,,
\end{eqnarray}
where $\lambda_1^{\gamma, 0}$ is real since the detailed balance condition holds.  
Thus the annealing schedule without satisfying the detailed balance condition may yield shorter calculation time with same calculation accuracy as one satisfying detailed balance condition: 
\begin{eqnarray}
\tau \le \tau_{\rm max}^{\gamma, u}\le \tau_{\rm max}^{\gamma, 0}\,.
\end{eqnarray}
The realization of such dynamics remains to be a future work.

\begin{acknowledgments}
A. Ichiki was supported by JSPS KAKENHI Grants No. JP17H06469. 
\end{acknowledgments}

\bibliography{calculation_time_ver2.bib}

\end{document}